\documentclass[aps,prb,reprint,superscriptaddress,nofootinbib]{revtex4-2}
\usepackage{graphicx,epsfig}
\usepackage{times}
\usepackage{graphics,dcolumn,bm,float}
\usepackage{amssymb,amsmath,rotate,color,dsfont}
\usepackage{xcolor}
\usepackage[breaklinks,colorlinks=true,urlcolor=blue,citecolor=red,linkcolor=blue]{hyperref}
\usepackage[T1]{fontenc}
\usepackage{soul}

\begin{document}
\unitlength 1 cm
\newcommand{\be}{\begin{equation}}
\newcommand{\ee}{\end{equation}}
\newcommand{\bearr}{\begin{eqnarray}}
\newcommand{\eearr}{\end{eqnarray}}
\newcommand{\nn}{\nonumber}
\newcommand{\vpdag}{{\vphantom{\dagger}}}
\newcommand{\vecr}{\vec{r}}
\newcommand{\bs}{\boldsymbol}
\newcommand{\up}{\uparrow}
\newcommand{\down}{\downarrow}
\newcommand{\fns}{\footnotesize}
\newcommand{\ns}{\normalsize}
\newcommand{\cdag}{c^{\dagger}}
\newcommand{\jh}{J_{\rm H}}
\newcommand{\tn}{T_{\rm N}}
\newcommand{\tq}{T_{q}}
\newcommand{\la}{\langle}
\newcommand{\ra}{\rangle}
\newcommand{\sgn}{\text{sgn}}
\newcommand{\inactiveref}[1]{\textcolor{blue}{\ref*{#1}}}

\definecolor{red}{rgb}{1.0,0.0,0.0}
\definecolor{green}{rgb}{0.0,1.0,0.0}
\definecolor{blue}{rgb}{0.0,0.0,1.0}

\newcommand{\red}[1]{\textcolor{red}{#1}}
\newcommand{\violet}[1]{\textcolor{violet}{#1}}
\newcommand{\blue}[1]{\textcolor{blue}{#1}}
\newcommand{\AQ}[1]{{\textcolor{red!80!black}{{[Alireza: #1]}}}}

\title{Altermagnetic Anomalous Hall Effect and Spin–Edge-Locked Chiral Modes\\in a Modified Kane–Mele–Hubbard Model}

\author{Mohsen Hafez-Torbati}
\email{m.hafeztorbati@gmail.com}
\affiliation{Department of Physics, Shahid Beheshti University, 1983969411, Tehran, Iran}

\author{Alireza Qaiumzadeh}
\email{alireza.qaiumzadeh@ntnu.no}
\affiliation{Center for Quantum Spintronics, Department of Physics,
Norwegian University of Science and Technology, NO-7491 Trondheim, Norway}

\begin{abstract}
We establish a correlation-driven route to the altermagnetic anomalous Hall effect (AHE) and its associated \emph{spin--edge-locked} edge states in a modified Kane--Mele--Hubbard model. Using dynamical mean-field theory (DMFT), we show that, at half-filling, increasing the Hubbard interaction drives the system from a metallic paramagnetic phase hosting antichiral edge states into an insulating in-plane N\'eel-type antiferromagnetic phase, in which a residual antiunitary symmetry forbids the AHE. Hole doping induces a spin-flop transition to an out-of-plane N\'eel-type antiferromagnetic phase, thereby breaking this symmetry and generating a finite anomalous Hall conductivity that persists into the strongly correlated regime. Distinct from a conventional spin-polarized Hall response in ferromagnets, the altermagnetic AHE receives equal and additive contributions from the two symmetry-related spin sectors and is accompanied by spin--edge-locked chiral states. Our results demonstrate that carrier doping and spin-rotationally invariant Hubbard interactions are sufficient to realize the altermagnetic AHE, without invoking an explicitly Ising-like interaction, and provide a realistic microscopic route toward its realization in correlated transition metal dichalcogenides monolayers.
\end{abstract}

\maketitle

{\it Introduction---}Altermagnets are a class of collinear antiferromagnetic (AF) materials that combine vanishing net magnetization with spin-split electronic bands without relying on typically weak relativistic spin-orbit coupling (SOC) \cite{noda2016momentum, hayami2019, naka2019spin, PhysRevB.99.184432, PhysRevB.102.014422,Smejkal2022b,Smejkal2020}. Conventional ferromagnets exhibit exchange-driven spin splitting but possess a finite net magnetization, whereas conventional collinear AFs have zero net magnetization but generally retain spin-degenerate bands because of the combined symmetry of time reversal with either spatial inversion $\mathcal{PT}$ or a lattice translation $\boldsymbol{t}\mathcal{T}$. In altermagnets, these degeneracy-enforcing symmetries are absent, while the opposite-spin sublattices are related by a lattice rotation. This distinctive symmetry produces momentum-dependent spin splitting with alternating spin polarization across the Brillouin zone. Considerable theoretical and experimental effort has therefore focused on identifying altermagnetic materials \cite{Smejkal2022a,Song2025}, culminating in the experimental observation of altermagnetic spin-split bands in $\alpha$-MnTe \cite{Krempasky2024,Lee2024} and CrSb \cite{Reimers2024}.

Altermagnets are expected to exhibit phenomena absent in conventional collinear antiferromagnets, most notably the anomalous Hall effect (AHE) \cite{Smejkal2020,Smejkal2022a}. The AHE is commonly observed in metallic ferromagnets \cite{Nagaosa2010} and antiferromagnets with noncollinear magnetic order \cite{Smejkal2022}. In conventional collinear AFs, the combined $\mathcal{PT}$ or $\boldsymbol{t}\mathcal{T}$ symmetry forbids a net intrinsic AHE. In altermagnets, the absence of these symmetries allows a nonzero momentum-integrated Berry curvature and, consequently, a finite AHE. Experimental signatures of the altermagnetic AHE have been reported in $\alpha$-MnTe \cite{Betancourt2023}, Mn$_5$Si$_3$ \cite{Reichlova2024}, and RuO$_2$ \cite{Feng2022}. For RuO$_2$, however, attributing the observed AHE to altermagnetism remains controversial, as recent studies have found no evidence for the proposed altermagnetic order \cite{alaei2026complexmagneticbehaviorruo2}.

In recent years, considerable attention has been devoted to understanding the microscopic origins of altermagnetism and its associated phenomena through the development and analysis of interacting lattice models \cite{Das2024,Sato2024,Giuli2025,Zhao2025,Wiedmann2026,Lin2026,Sato2026}. In particular, an altermagnetic Kane–Mele model with an antiferromagnetic Ising-like interaction was shown to exhibit an interaction-driven altermagnetic AHE \cite{Sato2024}. Although SOC is not required for the nonrelativistic spin splitting characteristic of altermagnets, it is required for a finite intrinsic AHE in this class of collinear systems. The Ising-like interaction explicitly favors easy-axis N\'eel-type AF ordering perpendicular to the honeycomb plane, denoted as the $z$-AF phase, over the in-plane  N\'eel-type AF ordering, denoted as the $xy$-AF phase. The $z$-AF phase supports a finite AHE, whereas the $xy$-AF phase retains a symmetry that forces the Hall response to vanish. Consequently, the previously reported altermagnetic AHE relies crucially on the strongly anisotropic Ising-like interaction. This raises the important question of whether an altermagnetic AHE can also emerge from a more realistic, spin-rotationally invariant electronic interaction, such as the Hubbard interaction.

Here, we investigate the altermagnetic Kane--Mele--Hubbard (AKMH) model using DMFT \cite{Georges1996}. At half-filling, increasing the Hubbard interaction drives a transition from a metallic phase hosting antichiral edge states to an $xy$-AF insulating phase. Unlike conventional chiral edge states, antichiral modes within each spin sector co-propagate along opposite edges and are compensated by counterpropagating bulk states \cite{Colomes2018}.
The two spin sectors are related by time-reversal symmetry and, in contrast to the quantum spin Hall state, the AHE is zero for each spin sector individually.

Upon doping, we uncover a spin-flop transition from the $xy$-AF phase to the $z$-AF phase, which enables the emergence of the altermagnetic AHE. By demonstrating that this effect can arise from the spin-rotationally invariant Hubbard interaction, our results provide a realistic microscopic foundation for the proposed realization of the altermagnetic AHE in transition-metal dichalcogenide (TMD) monolayers \cite{Sato2024}. We further show that a sizable anomalous Hall conductivity is accompanied by \emph{spin--edge-locked} chiral
modes involving both spin sectors, reflecting their simultaneous contributions to the altermagnetic AHE.

{\it Altermagnetic Kane--Mele--Hubbard model---}The altermagnetic Kane–Mele model preserves time-reversal symmetry but breaks the spatial-inversion symmetry of the conventional Kane–Mele model, thereby allowing non-relativistic spin-split energy bands at generic momenta \cite{Sato2024}. The AKMH Hamiltonian is given by
\begin{align}
H=&-t\sum_{\langle i,j\rangle} \sum_{\alpha} c^{\dag}_{i\alpha} c^{\vpdag}_{j\alpha}
+{\rm i}\lambda \sum_{\langle\langle i,j \rangle\rangle}
\sum_{\alpha \beta} \xi_i^{\vpdag}
\nu_{ij}^{\vpdag}
c^{\dag}_{i\alpha} \sigma^{z}_{\alpha\beta} c^{\vpdag}_{j\beta} \nn \\
&+U \sum_{i} n_{i,\up} n_{i,\down} -\mu \sum_{i,\alpha} n_{i,\alpha} \ ,
\label{eq:akmh}
\end{align}
where $c_{i\alpha}^{\dagger}$ ($c_{i\alpha}$) creates (annihilates) an electron with spin projection $\alpha=\uparrow,\downarrow$ at site $i$. The sum over $i$ runs over the honeycomb lattice, which lies in the $xy$ plane. The symbols $\langle i,j\rangle$ and $\langle\langle i,j \rangle\rangle$ denote nearest-neighbor (NN) and next-nearest-neighbor (NNN) pairs, respectively. Furthermore, $\sigma^z$ is the Pauli matrix in spin space, and $\nu_{ij}=(2/\sqrt{3})(\hat{d}_1\times\hat{d}_2)_z=\pm1$, where $\hat{d}_1$ and $\hat{d}_2$ are unit vectors along the two consecutive bonds connecting site $j$ to site $i$. The first two terms define the altermagnetic Kane--Mele model. Its distinction from the conventional Kane--Mele model \cite{Kane2005b} lies in the sublattice-dependent factor $\xi_i=\pm1$ in the second term, which takes opposite signs on the two sublattices and thereby breaks spatial-inversion symmetry. This staggered SOC term has been proposed as a minimal description of inversion-asymmetric transition-metal dichalcogenide monolayers \cite{Colomes2018,Xiao2012}. Throughout this work, we set the SOC strength to $\lambda=0.1t$.
The third term in Eq.~\eqref{eq:akmh} describes the onsite Hubbard interaction, where $n_{i\alpha}:=c^{\dagger}_{i\alpha}c_{i\alpha}$ is the occupation-number operator. The last term represents the chemical potential, which is adjusted to obtain the desired filling. Owing to the particle-hole symmetry of the model, half-filling corresponds to $\mu=U/2$.

\begin{figure}[t]
   \begin{center}
   \includegraphics[width=0.35\textwidth,angle=0]{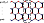}
   \caption{Honeycomb structure with periodic boundary condition in $x$
   and open periodic boundary condition in $y$ direction. The number of sites
   in $y$ direction is $N_y$. The zigzag edges at $y=0$ and at $y=N_y-1$ are emboldened.
   The dashed box denotes the unit cell.}
   \label{fig:lat}
   \end{center}
\end{figure}

DMFT is a well-established method for studying strongly correlated systems with local electron-electron interactions \cite{Georges1996}. Unlike static mean-field theory, DMFT retains the full frequency dependence of the self-energy while approximating it as spatially local and, hence, momentum independent. It therefore fully captures local quantum fluctuations while neglecting nonlocal ones. Our analysis of the model in Eq.~\eqref{eq:akmh} is based on real-space DMFT \cite{Potthoff1999,Song2008,Snoek2008}. Specifically, we employ the implementation introduced in Ref.~\cite{Hafez-Torbati2018}, which accommodates different types of magnetic order and provides access to both bulk and edge properties within the same framework.

Within this approximation, the self-energy takes the form $\Sigma_{i\alpha,j\beta}({\rm i}\omega_n)=\delta_{ij}\Sigma_{i;\alpha\beta}({\rm i}\omega_n)$. The off-diagonal spin components, $\alpha\neq\beta$, are retained and become nonzero in the $xy$-AF phase. The lattice model is mapped onto a set of Anderson impurity models, whose number is determined by the number of inequivalent lattice sites in the chosen unit cell. To investigate the bulk properties, we impose periodic boundary conditions along both spatial directions, resulting in two impurity models corresponding to the two sublattices. To study the edge properties, we instead consider the cylindrical geometry illustrated in Fig.~\ref{fig:lat}, with periodic boundary conditions along $x$ and zigzag edges at $y=0$ and $y=N_y-1$. The resulting system is described by $2N_y$ impurity models.

We set the temperature to $T=0.05t$ and solved the impurity models using exact diagonalization (ED) \cite{Georges1996,Caffarel1994}. ED-based DMFT has been successfully applied to several closely related models \cite{Ebrahimkhas2021,Ebrahimkhas2022,Vanhala2016}, yielding results in good agreement with those obtained using other numerical methods, including the density-matrix renormalization group \cite{He2024}. Furthermore, nonlocal quantum fluctuations have been shown to introduce only minor quantitative corrections in comparable systems \cite{Mertz2019}.

The anomalous Hall conductivity vanishes identically in both the paramagnetic and $xy$-AF phases. In the paramagnetic phase, this follows directly from time-reversal symmetry. Although time-reversal symmetry is broken in the $xy$-AF phase, the phase remains invariant under the combined antiunitary symmetry $\mathcal{C}_{2z}^{s}\mathcal{T}$, where $\mathcal{C}_{2z}^{s}$ denotes a $\pi$ rotation of the spin about the $z$ axis. This symmetry reverses the Berry curvature, $\Omega(\vec{k})=-\Omega(-\vec{k})$, and therefore forces its Brillouin-zone integral, and hence the Hall conductivity, to vanish. By contrast, the $z$-AF phase breaks this combined symmetry and can support a finite Hall response. We compute its Hall conductivity from the imaginary-frequency Green's function using the Ishikawa--Matsuyama formula \cite{Ishikawa1987},
\begin{align}
\label{eq:hall}
\sigma_{xy}=\frac{e^2}{h}\frac{T}{12\pi} \epsilon_{\mu\nu\rho}~
{\rm Im}\!\left[ \sum_{\alpha} \sum_n \int \!{\rm d}\vec{k}
~{\rm Tr}\!\left[
\boldsymbol{G}_\alpha \partial_\mu \boldsymbol{G}_\alpha^{-1}
\right.
\right. \nn \\
\left.
\left.
\times
\boldsymbol{G}_\alpha \partial_\nu \boldsymbol{G}_\alpha^{-1}
\boldsymbol{G}_\alpha \partial_\rho \boldsymbol{G}_\alpha^{-1}
\right]
\vphantom{\sum_\alpha\int}
\right] \ ,
\end{align}
where $\boldsymbol{G}^{\vpdag}_\alpha \!\equiv\! \boldsymbol{G}^{\vpdag}_\alpha({\rm i}\omega_n,\vec{k})\!=\!
\left[ {\rm i}\omega_n \boldsymbol{\mathds{1}}-\boldsymbol{\mathcal{H}}^{(0)}_\alpha(\vec{k})-\boldsymbol{\Sigma}^{\vpdag}_\alpha({\rm i}\omega_n) \right]^{-1}$,
$\boldsymbol{\mathcal{H}}^{(0)}_\alpha(\vec{k})$ is the Bloch Hamiltonian matrix of Eq. \eqref{eq:akmh},
$\boldsymbol{\Sigma}^{\vpdag}_\alpha({\rm i}\omega_n)$ stands for the self-energy in the
DMFT approximation,
$\epsilon_{\mu\nu\rho}$ is the totally antisymmetric tensor, and summations are assumed over the indices
$\mu$, $\nu$, and $\rho$, each of which runs over $\mathrm{i}\omega_n$, $k_x$, and $k_y$. Further details for the
computation of the Hall conductivity from the relation \eqref{eq:hall} can be found in Ref. \cite{HafezTorbati2026}.
Notably, our calculation of the altermagnetic anomalous Hall conductivity incorporates dynamical electronic correlations and thus goes beyond the static mean-field treatment employed in the previous analysis \cite{Sato2024}.

To characterize the single-particle excitation spectrum, we compute the single-particle spectral function from the real-frequency Green's function,
\be
A_{\vec{d},\alpha}(\omega,\vec{k})=
-{\frac{1}{\pi}\rm Im}\! \left[\boldsymbol{G}(\omega+{\rm i}\eta,\vec{k})\right]_{\vec{d}\alpha,\vec{d}\alpha}
\label{eq:spectral}
\ee
where $\vec{d}$ denotes a lattice site in the unit cell and $\eta=0.01t$ is the broadening parameter. Equation~\eqref{eq:spectral} also applies to the $xy$-AF phase, although the $z$ component of spin is not conserved, and $\alpha$ therefore denotes only the spin projection in the chosen basis. The local spectral function $A_{\vec{d},\alpha}(\omega)$ is obtained by integrating the momentum-resolved spectral function over the Brillouin zone, with the normalization chosen to satisfy the spectral sum rule. The Bloch Hamiltonian for the cylindrical geometry with zigzag edges, illustrated in Fig.~\ref{fig:lat}, is provided in Supplemental Materials \cite{sm}.

{\it Magnetic phase transition at half-filling---}The altermagnetic Kane--Mele model at half-filling hosts a metallic paramagnetic phase with antichiral edge states \cite{Colomes2018}. For a given spin projection, the edge modes propagate in the same direction along the two parallel edges and are compensated by counterpropagating bulk modes. The opposite-spin modes are related by time-reversal symmetry and propagate in opposite directions.
Upon increasing the Hubbard interaction in the AKMH model, Eq.~\eqref{eq:akmh}, the metallic paramagnetic phase undergoes a transition to an insulating AF phase. The orientation of the local magnetic moments, either perpendicular to the honeycomb plane in the $z$-AF phase or within the plane in the $xy$-AF phase, is crucial for the emergence of the altermagnetic AHE. As we discussed before, the AHE vanishes in the $xy$-AF phase because it preserves the combined antiunitary symmetry $\mathcal{C}_{2z}^{s}\mathcal{T}$, which forces the momentum-integrated Berry curvature to vanish, whereas the $z$-AF phase breaks this symmetry and can therefore support a finite AHE.

\begin{figure}[t]
   \begin{center}
   \includegraphics[width=0.24\textwidth,angle=-90]{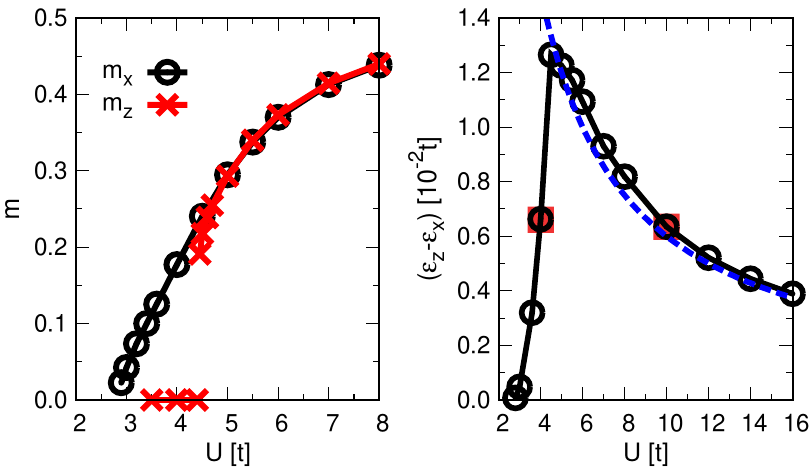}
   \caption{(a) Local magnetization for the $z$-AF solution ($m_z$) and for the $xy$-AF solution ($m_x$) versus the Hubbard interaction $U$.
   (b) The energy difference between the $z$-AF solution and the $xy$-AF solution per lattice site versus $U$.
   The blue dashed line displays the mean-field results of the effective low-energy spin model valid at the large-$U$ limit, see the main text.
   Note that the $z$-AF solution does not exist below $U \approx 4.5t$ and $\varepsilon_z$ denotes the energy per lattice site of the paramagnetic
   solution instead. The spin-orbit coupling is fixed to $\lambda=0.1t$. The results are for the number of bath sites
   $n_b=6$ in the ED impurity solver except for the filled red squares at $U=4t$ and $10t$ which are for $n_b=7$.}
   \label{fig:mag}
   \end{center}
\end{figure}

Figure~\ref{fig:mag}(a) compares the local sublattice magnetizations of the $z$-AF and $xy$-AF solutions,
denoted by $m_z$ and $m_x$, respectively, while Fig.~\ref{fig:mag}(b) shows their energy difference as a function of $U$.
The results were obtained using $n_b=6$ bath sites, except for the red filled squares in Fig.~\ref{fig:mag}(b), which correspond to $n_b=7$. The results for $n_b=5$, not shown, also agree closely with those for $n_b=6$ and $7$. As shown in Fig.~\ref{fig:mag}(a), the $xy$-AF solution persists down to smaller values of $U$ than the $z$-AF solution. Accordingly, for $U<4.5t$, Fig.~\ref{fig:mag}(b) compares the energies of the paramagnetic and $xy$-AF solutions. The blue dashed line shows the mean-field prediction, $\varepsilon_z-\varepsilon_{xy}=Z_2\lambda^2/U$, obtained from the effective low-energy spin model \cite{Ebrahimkhas2022},
\be
\label{eq:spin}
H_{\rm eff}=J_1\sum_{\langle i,j \rangle} \vec{S}_i \cdot \vec{S}_j
+J_2\sum_{\langle\langle i,j \rangle\rangle} ( {S}_i^z {S}_j^z - {S}_i^x {S}_j^x -{S}_i^y {S}_j^y )
\ee
valid in the large-$U$ limit with $J_1=4t^2/U$ and $J_2=4\lambda^2/U$.
The NNN coordination number $Z_2=6$ holds.
In writing Eq.~\eqref{eq:spin}, each lattice bond is counted only once. The anisotropic NNN interaction, which is ferromagnetic in the $xy$ plane and antiferromagnetic along the $z$ direction, favors N\'eel order with the local sublattice magnetization oriented within the $xy$ plane rather than along the $z$ direction. The results shown in Fig.~\ref{fig:mag} confirm that, at half-filling, the AKMH model undergoes a transition from the paramagnetic phase to the $xy$-AF phase as the Hubbard interaction increases. The $xy$-AF phase is energetically favored over the $z$-AF phase, with their energy difference approaching the prediction of the effective spin model in the large-$U$ limit \cite{Hafez-Torbati2022}.

The charge excitations of the system at half-filling are addressed in Supplemental Materials \cite{sm}.
The bulk spectral function and the charge gap versus the Hubbard $U$ indicate
gapless excitations in the paramagnetic phase which become gapped as the system enters the $xy$-AF phase.
Therefore, the magnetic ordering at half-filling is accompanied by a metal--insulator transition.
Furthermore, the momentum-resolved spectral function for the cylindrical geometry unveils that
the antichiral edge modes proposed for the non-interacting model \cite{Colomes2018} persists to the finite
interaction strengths in the metallic paramagnetic state \cite{sm}.

\begin{figure}[t]
   \begin{center}
   \includegraphics[width=0.277\textwidth,angle=-90]{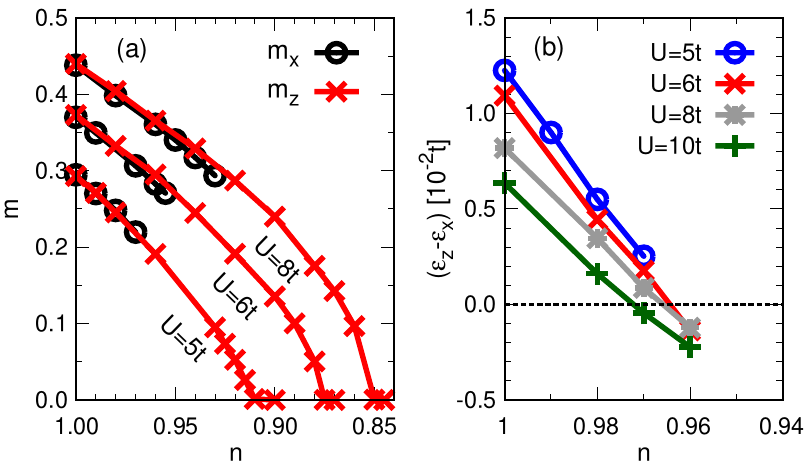}
   \caption{(a) Local magnetization for the $z$-AF solution ($m_z$) and for the $xy$-AF solution ($m_x$) versus
   the electron density $n$ for different values of the Hubbard interaction $U$.
   (b) The energy difference per lattice site between the $z$-AF solution and the $xy$-AF solution
   versus $n$ for different values of $U$. The data are obtained for the model parameter $\lambda=0.1t$, the temperature
   $T=0.05t$, and the number of bath sites $n_b=6$ in the ED impurity solver.}
   \label{fig:doped:mag}
   \end{center}
\end{figure}

{\it Doping-induced spin-flop transition and altermagnetic AHE---}At half-filling, the $xy$-AF solution is always energetically favored over the $z$-AF solution, as shown in Fig.~\ref{fig:mag}. Away from half-filling, however, we find a doping-induced spin-flop transition from the $xy$-AF phase to the $z$-AF phase. Figure~\ref{fig:doped:mag}(a) shows the N\'eel order parameters $m_z$ and $m_x$ as functions of the electron density $n$ for several values of $U$. As $n$ decreases, $m_z$ decreases continuously and eventually vanishes. By contrast, the $xy$-AF solution ceases to satisfy the self-consistent DMFT equations below a $U$-dependent density, accounting for the absence of $m_x$ data at lower values of $n$.

\begin{figure}[t]
   \begin{center}
   \includegraphics[width=0.25\textwidth,angle=-90]{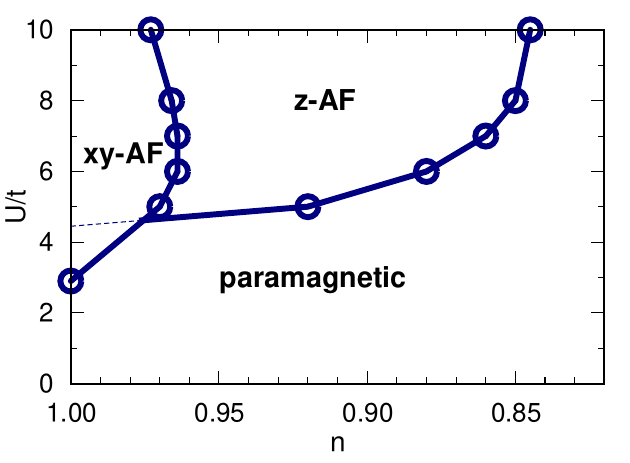}
   \caption{Magnetic phase diagram of the altermagnetic Kane-Mele-Hubbard model in Eq. \eqref{eq:akmh}.
   The horizontal axis represents the electron density $n$ and the vertical axis represents the Hubbard
   interaction $U$. The paramagnetic, the $xy$-AF, and the $z$-AF phases are distinguished.
   The thin dashed line connects the $z$-AF phase boundary to its ending point $U \approx 4.5t$ at half-filling.
   The spin-orbit coupling is fixed to $\lambda=0.1t$ and the temperature to $T=0.05t$.
   The number of bath sites $n_b=6$ is used in the ED impurity solver.}
   \label{fig:pd}
   \end{center}
\end{figure}

Figure~\ref{fig:doped:mag}(b) shows the energy difference per lattice site between the two AF solutions. For all interaction strengths considered except $U=5t$, the $z$-AF solution becomes energetically favorable before the metastable $xy$-AF solution disappears, demonstrating that the spin-flop transition is driven by an energetic crossing rather than by the loss of the $xy$-AF solution. For $U\gtrsim6t$, the transition shifts toward lower electron densities as $U$ increases. The resulting magnetic phase diagram in the $n$--$U$ plane is summarized in Fig.~\ref{fig:pd}.

\begin{figure}[b]
   \begin{center}
   \includegraphics[width=0.3\textwidth,angle=-90]{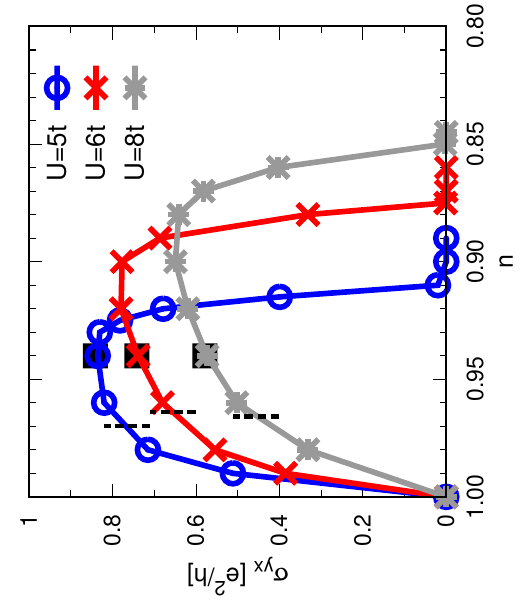}
   \caption{The anomalous Hall conductivity $\sigma_{yx}=-\sigma_{xy}$ for the $z$-AF solution
   plotted versus the electron density $n$ at different values of the Hubbard interaction $U$.
   The vertical black dashed line for each value of $U$ specifies the electron density above
   which the system becomes a $xy$-AF phase.
   The anomalous Hall conductivity is entirely zero in the $xy$-AF phase.
   The spin-orbit coupling is fixed to $\lambda=0.1t$ and the temperature to $T=0.05t$.
   The results are for the number of bath sites $n_b=6$ except for the filled black squares at
   $n=0.94$ and different values of $U$, which are for $n_b=8$.}
   \label{fig:ahe}
   \end{center}
\end{figure}

Within the present model, stabilization of the $z$-AF phase is a prerequisite for the emergence of the altermagnetic AHE. The anomalous Hall conductivity vanishes identically in the $xy$-AF phase owing to the residual antiunitary symmetry $\mathcal{C}_{2z}^{s}\mathcal{T}$, which forces the momentum-integrated Berry curvature to vanish. By contrast, the $z$-AF phase breaks this symmetry and can support a finite anomalous Hall conductivity. Figure~\ref{fig:ahe} shows $\sigma_{yx}=-\sigma_{xy}$, calculated for the $z$-AF solution using Eq.~\eqref{eq:hall}, as a function of the electron density $n$ for several values of $U$. The results were obtained using $n_b=6$ bath sites, except for the filled black squares at $n=0.94$, which correspond to $n_b=8$. The results for $n_b=7$ also agree closely with those obtained using $n_b=6$ and $8$. Upon doping away from half-filling, $\sigma_{yx}$ becomes finite, reaches a maximum magnitude, and subsequently vanishes at the transition to the paramagnetic phase. At a fixed electron density, its magnitude generally decreases with increasing $U$, although a finite Hall response persists even at the largest interaction strengths considered.

By incorporating dynamical electronic correlations and local quantum fluctuations, our calculation of the altermagnetic AHE goes beyond the static mean-field treatment of Ref.~\cite{Sato2024}. Moreover, demonstrating that the AHE can emerge from the spin-rotationally invariant Hubbard interaction, rather than requiring an explicitly Ising-like interaction, provides a more realistic microscopic foundation for the proposed realization of the altermagnetic AHE in TMD monolayers \cite{Sato2024}.

In contrast to a conventional spin-polarized AHE dominated by a single spin sector, both spin sectors contribute equally and with the same sign to the altermagnetic anomalous Hall conductivity shown in Fig.~\ref{fig:ahe}. This equality follows from the symmetry that relates the two spin sectors through a combined spin reversal and crystalline transformation, while allowing their transverse charge responses to add rather than cancel.

A quantum AHE has also been reported in other unconventional collinear AF systems \cite{Jiang2018,Ebrahimkhas2022,HafezTorbati2026,Hafez-Torbati2024,Liu2024}.
In those phases, however, no antiunitary symmetry combining time reversal with a space-group operation restores the magnetic structure. Their Hall response is therefore analogous to that of a time-reversal-broken magnetic state and, in the models considered in Refs.~\cite{Jiang2018,Ebrahimkhas2022,HafezTorbati2026,Hafez-Torbati2024,Liu2024}, is dominated by a single spin sector. By contrast, the Hall response shown in Fig.~\ref{fig:ahe} receives equal, additive contributions from the two symmetry-related spin sectors, providing a distinct signature of its altermagnetic origin.

\begin{figure}[t]
   \begin{center}
   \includegraphics[width=0.35\textwidth,angle=0]{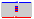}
   \caption{Schematic representation of \emph{spin--edge-locked} chiral modes accompanying the altermagnetic AHE. Blue and red denote spin-$\up$ and spin-$\down$, respectively. The two parallel edges host counterpropagating modes with opposite spin polarizations, while the edge current within each spin sector is compensated by counterpropagating bulk states. Unlike conventional chiral edge modes in a Chern insulator, these modes occur in a metallic phase with a nonquantized Hall conductivity and are distinguished by their edge-dependent spin polarization.}
   \label{fig:edge:af:schem}
   \end{center}
\end{figure}

{\it Spin--edge-locked chiral modes---}In this section, we show that the altermagnetic AHE is accompanied by spin--edge-locked chiral modes involving both spin sectors. As schematically illustrated in Fig.~\ref{fig:edge:af:schem}, the two parallel edges host counterpropagating modes with opposite spin polarizations. Thus, although each edge mode has a definite spin projection, the complete chiral edge structure involves both spin sectors. This behavior reflects the equal and additive contributions of the two spin sectors to the altermagnetic AHE.

\begin{figure}[t]
   \begin{center}
   \includegraphics[width=0.47\textwidth,angle=0]{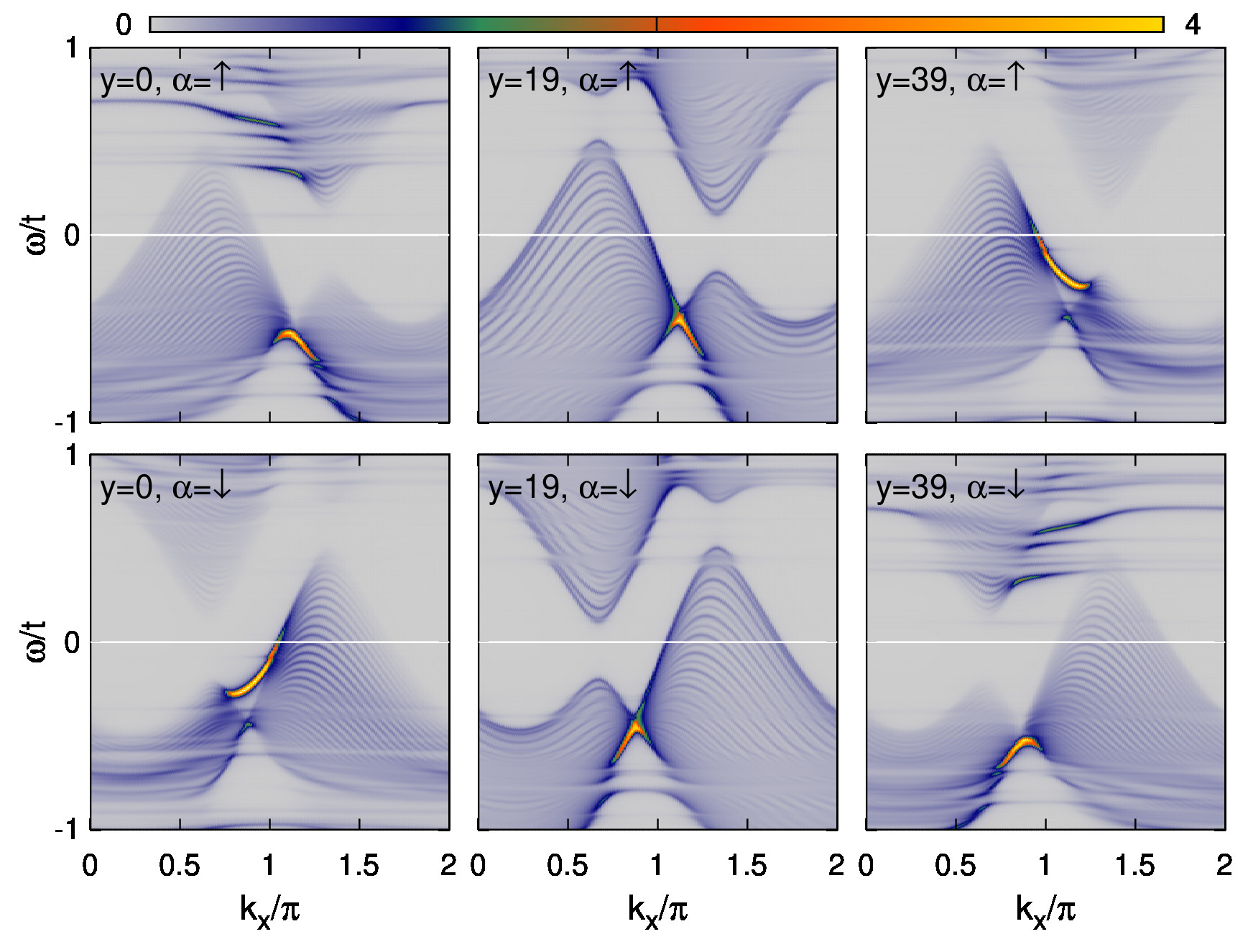}
   \caption{Spectral function $A_{y,\alpha}(\omega,k_x)$ averaged over the two nonequivalent lattice sites
   in the $x$ direction for the cylindrical geometry sketched in Fig. \ref{fig:lat} with $N_y=40$.
   The results for spin $\up$ and $\down$ at the lower edge $y=0$, the bulk $y=19$, and the upper edge $y=39$
   are provided. The horizontal white line at $\omega=0$ specifies the Fermi energy.
   The results are for the spin-orbit coupling $\lambda=0.1t$ and the Hubbard interaction $U=5.0t$ at the electron density $n=0.93$ in the $z$-AF phase.
   The temperature is set to $T=0.05t$
   and the number of bath sites $n_b=6$ is used in the ED impurity solver.}
   \label{fig:edge:af}
   \end{center}
\end{figure}

Figure~\ref{fig:edge:af} shows the momentum-resolved spectral function $A_{y,\alpha}(\omega,k_x)$ for the cylindrical geometry illustrated in Fig.~\ref{fig:lat}, with $N_y=40$. The results are obtained in the $z$-AF phase at $U=5t$ and electron density $n=0.93$, where the anomalous Hall conductivity is close to its maximum, as shown in Fig.~\ref{fig:ahe}. The calculations use $n_b=6$ bath sites in the ED impurity solver. The horizontal white line at $\omega=0$ marks the Fermi energy. A spin-$\up$ edge mode is visible at the upper edge, $y=39$, while a counterpropagating spin-$\down$ edge mode appears at the lower edge, $y=0$. These edge-localized modes are absent from the representative bulk spectrum at $y=19$. The spin--edge-locked chiral modes shown in Fig.~\ref{fig:edge:af} evolve from the antichiral edge modes of the paramagnetic phase [Fig.~\textcolor{blue}{2} in Supplemental Materials]
upon the development of $z$-AF order and the concomitant emergence of the altermagnetic AHE.

As the Hubbard interaction increases, the spin--edge-locked chiral modes disappear together with the suppression of the anomalous Hall conductivity. If an altermagnetic quantum anomalous Hall phase is realized, it should host robust spin--edge-locked chiral modes involving both spin sectors over a finite parameter range. Unlike the metallic AHE phase studied here, such a phase would possess a bulk insulating gap and therefore lack compensating bulk states at the Fermi energy. Spin-mixing terms may be necessary to open a topologically nontrivial bulk gap and stabilize this phase, although this possibility lies beyond the spin-conserving model considered here. We emphasize that the (quantum) AHE in altermagnets can differ fundamentally from its conventional ferromagnetic counterpart: both symmetry-related spin sectors contribute equally and additively to the Hall response. This distinctive behavior originates from the crystalline symmetry relating the two spin sectors of the altermagnetic phase.

{\it Concluding remarks---}Our results reveal doping as a means of controlling not only the strength of AF order but also its orientation and the associated transverse response in the AKMH model. At half-filling, the anisotropic NNN exchange selects the $xy$-AF phase, whose residual antiunitary symmetry precludes an AHE. Hole doping reverses this magnetic anisotropy and stabilizes the $z$-AF phase, thereby enabling a sizable AHE. This establishes the doping-induced spin-flop transition as the key link between Hubbard correlations and altermagnetic Hall transport.

The resulting Hall response has a distinctive boundary manifestation: the antichiral edge states of the paramagnetic metal evolve into spin--edge-locked chiral  modes involving both spin sectors upon the onset of $z$-AF order. Their equal and additive contributions distinguish this response from a conventional spin-polarized AHE. More broadly, our findings show that neither an explicitly Ising-like interaction nor a built-in out-of-plane magnetic anisotropy is required to realize the altermagnetic AHE; the required magnetic state can instead emerge through the interplay of carrier doping, SOC, and Hubbard correlations. This mechanism strengthens the microscopic basis for pursuing electrically tunable altermagnetic Hall transport in correlated TMD monolayers. Extending the model to include spin-mixing terms may further stabilize a bulk-insulating phase with robust chiral boundary transport, providing a possible route toward an altermagnetic quantum anomalous Hall state.

{\it{Acknowledgments---}}
 A.Q. was supported by the Research Council of Norway through Grant Nos. 353919 and 361800 ``QTransMag'', and Grant No. 262633 ``QuSpin''.

{\it{Data Availability---}}
The numerical data supporting the findings of this study are available from the corresponding authors upon reasonable request. 


%

\end{document}


\newcommand{\be}{\begin{equation}}
\newcommand{\ee}{\end{equation}}
\newcommand{\bearr}{\begin{eqnarray}}
\newcommand{\eearr}{\end{eqnarray}}
\newcommand{\bseq}{\begin{subequations}}
\newcommand{\eseq}{\end{subequations}}
\newcommand{\nn}{\nonumber}
\newcommand{\dagg}{{\dagger}}
\newcommand{\vpdag}{{\vphantom{\dagger}}}
\newcommand{\vecr}{\vec{r}}
\newcommand{\bs}{\boldsymbol}
\newcommand{\up}{\uparrow}
\newcommand{\down}{\downarrow}
\newcommand{\fns}{\footnotesize}
\newcommand{\ns}{\normalsize}
\newcommand{\cdag}{c^{\dagger}}
\newcommand{\so}{\lambda_{\rm SO}}
\newcommand{\jh}{J_{\rm H}}
\newcommand{\tn}{T_{\rm N}}
\newcommand{\tq}{T_{q}}
\newcommand{\la}{\langle}
\newcommand{\ra}{\rangle}
\newcommand{\sgn}{\text{sgn}}

\title{Supplemental Material:\\ Altermagnetic Anomalous Hall Effect and Spin–Edge-Locked Chiral Modes\\in a Modified Kane–Mele–Hubbard Model}

\author{Mohsen Hafez-Torbati}
\email{m.hafeztorbati@gmail.com}
\affiliation{Department of Physics, Shahid Beheshti University, 1983969411, Tehran, Iran}

\author{Alireza Qaiumzadeh}
\email{alireza.qaiumzadeh@ntnu.no}
\affiliation{Center for Quantum Spintronics, Department of Physics,
Norwegian University of Science and Technology, NO-7491 Trondheim, Norway}

\maketitle

\section{Bloch Hamiltonian for cylindrical geometry with zigzag edges}
To study the edge states in the altermagnetic Kane--Mele--Hubbard (AKMH) model we consider the cylindrical geometry with the zigzag edges as
sketched in Fig. 1 in the main text. There are open boundary conditions in $y$
and periodic boundary conditions in $x$ direction. We treat the honeycomb structure as a brick wall, where
each lattice site in the unit cell is specified by $(x,y)$ with $x=0,1$ and $y=0,1,\cdots,N_y-1$. After a
Fourier transform in the $x$ direction we obtain the Bloch Hamiltonian
\begin{subequations}
\label{eq:Hk}
\begin{gather}
 H^{(0)}(k_x)=H_{t}(k_x)+H_{\rm so}(k_x)+H_{\mu}(k_x) \ , \\
 H_{t}(k_x)=-t\sum_{\alpha=\up,\down} \left( \sum_{y=1,3}^{N_y-1} c^{\dagg}_{0,y+1;\alpha} c^{\vpdag}_{0,y;\alpha}
 +\sum_{y=0,2}^{N_y-1} c^{\dagg}_{1,y+1;\alpha} c^{\vpdag}_{1,y;\alpha}
 + 2\cos(k_x a) \sum_{y=0,1}^{N_y-1} c^{\dagg}_{1,y;\alpha} c^{\vpdag}_{0,y;\alpha}
 \right) +{\rm H.c.} \ , \\
 H_{\rm so}(k_x)=2\lambda \!\! \sum_{\alpha=\up,\down} ~\sum_{y=0,1}^{N_y-1} ~ \sum_{x=0}^{1} \sigma_{\!\alpha \alpha}^z
 \left\{ \sin(2k_x a) c^{\dagg}_{x,y;\alpha} c^{\vpdag}_{x,y;\alpha} -\sin(k_x a) \left(c^{\dagg}_{x+1,y+1;\alpha} c^{\vpdag}_{x,y;\alpha}
 +{\rm H.c.} \right) \right\} \ , \\
 H_{\mu}(k_x)=-\mu \sum_{\alpha=\up,\down} ~ \sum_{y=0,1}^{N_y-1} ~ \sum_{x=0}^1
 c^{\dagg}_{x,y;\alpha} c^{\vpdag}_{x,y;\alpha} \quad.
\end{gather}
\end{subequations}
The dependence of $c^{\dagg}_{x,y;\alpha}$ on the momentum $k_x$ is implicit,
$c^{\dagg}_{2,y;\alpha}\equiv c^{\dagg}_{0,y;\alpha}$, and $c^{\dagg}_{x,N_y;\alpha}\equiv 0$. The parameter $a$
represents the distance between the NN sites.
Note that the distance $2a$ is used as the unit of length in Fig. 7 in the main text and also in Fig. \ref{fig:edge} below.
The Bloch Hamiltonian \eqref{eq:Hk} can be seen
as an effective two-leg ladder model involving hopping up to the next-nearest neighbor.

The  Green's function ${\bf G}(\omega+{\rm i}\eta,k_x)$ is found using
the matrix representation of the Bloch Hamiltonian in Eq. \eqref{eq:Hk} and the DMFT self-energy $\bs{\Sigma}(\omega+{\rm i}\eta)$.
The momentum-resolved spectral function $A_{\vec{d},\alpha}(\omega,k_x)$ for the spin component $\alpha$ at the lattice site
$\vec{d}=(x,y)$ in the unit cell is then obtained via Eq. 3 in the main text.
We have used the broadening factor $\eta=0.01t$ as mentioned in the main text.

\section{Metal-insulator transition at half-filling}
\begin{figure}[t]
   \begin{center}
   \includegraphics[width=0.36\textwidth,angle=-90]{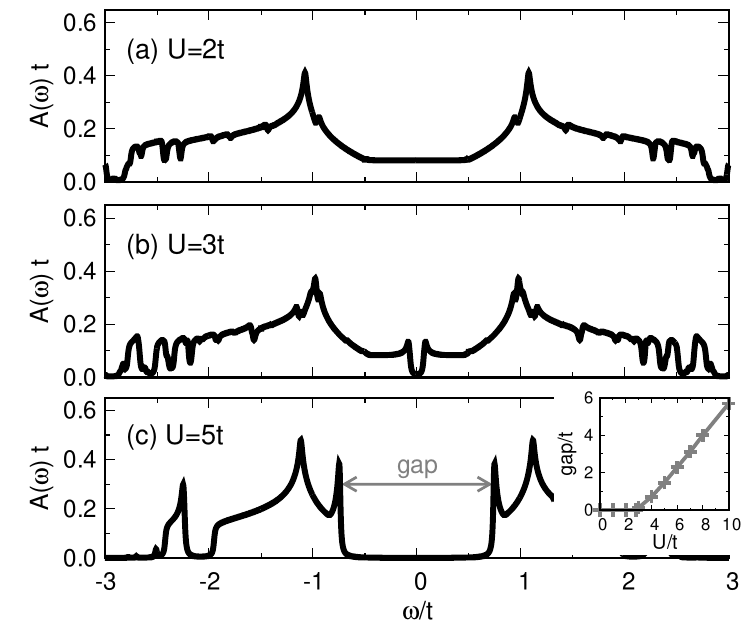}
   \caption{(a) Local spectral function averaged over the two sublattices and the spin for different values of Hubbard $U$ across the
   magnetic transition point at $U_c \approx 2.8t$. The inset represents the gap, as indicated in (c), vs $U$.
   The results are for the model parameter $\lambda=0.1t$, the temperature $T=0.05t$, and the number of bath sites
   $n_b=6$ in the ED impurity solver.}
   \label{fig:spectral}
   \end{center}
\end{figure}

To characterize the charge excitations at half-filling
across the magnetic transition at $U_c\approx2.8t$, we calculate the local spectral function averaged over the two sublattices and spin sectors for several values of $U$, as shown in Fig.~\ref{fig:spectral}. With increasing $U$, the finite spectral weight at the Fermi energy, $\omega=0$, is suppressed, and a spectral gap opens upon entering the $xy$-AF phase. The gap extracted from the spectrum, as indicated in Fig.~\ref{fig:spectral}(c), is plotted as a function of $U$ in the inset. The smaller critical interaction $U_c\approx2.8t$ compared with the estimate $U_c\approx3.9t$ for the half-filled honeycomb-lattice Hubbard model \cite{Sorella2012,Meng2010} can be partly attributed to their distinct noninteracting electronic structures. In particular, the AKMH model has a finite spectral weight at the Fermi energy, as shown in Fig.~\ref{fig:spectral}(a), whereas the honeycomb-lattice Hubbard model has a vanishing density of states at the Dirac points. In addition, the neglect of nonlocal quantum fluctuations within DMFT generally favors an earlier onset of magnetic order and may further underestimate the value of $U_c$.

\section{Antichiral edge states}
Antichiral edge states have been predicted in TMD monolayers and shown to be robust against disorder \cite{Colomes2018}. However, their stability against electron-electron interactions, which can be substantial in TMD monolayers, has not been explicitly investigated. For each spin sector, the antichiral edge modes propagate in the same direction along the two parallel edges, while the vanishing net current is ensured by counterpropagating bulk states.

\begin{figure}[htbp]
   \begin{center}
   \includegraphics[width=0.54\textwidth,angle=0]{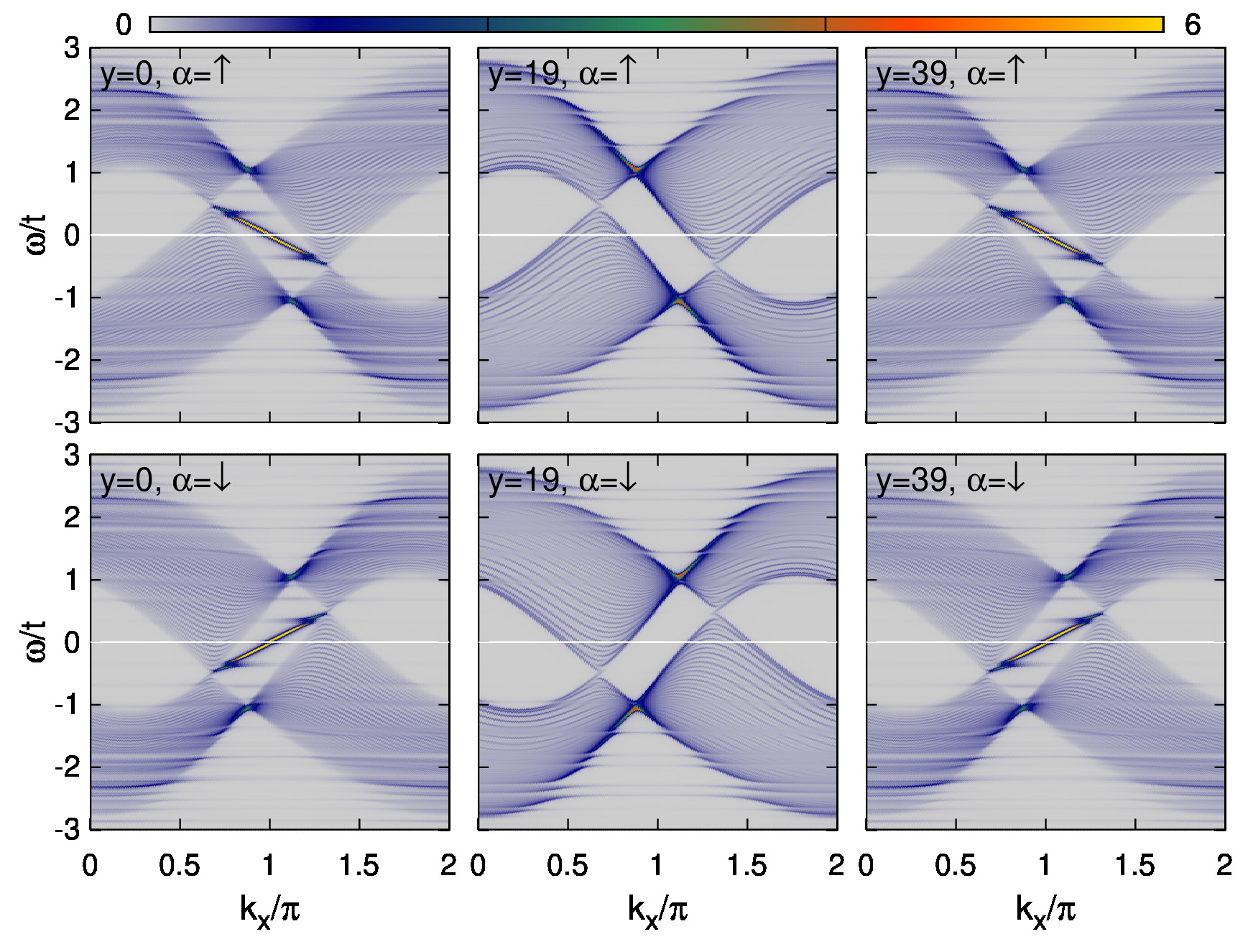}
   \caption{Spectral function $A_{y,\alpha}(\omega,k_x)$ averaged over the two nonequivalent lattice sites
   in the $x$ direction for the cylindrical geometry sketched in Fig. 1 in the main text with $N_y=40$.
   The results for spin $\up$ and $\down$ at the lower edge $y=0$, the bulk $y=19$, and the upper edge $y=39$
   are provided. The horizontal white line at $\omega=0$ specifies the Fermi energy.
   The results are for the spin-orbit coupling $\lambda=0.1t$ and the Hubbard interaction $U=2.0t$ at the half-filling
   $n=1$. The temperature is set to $T=0.05t$
   and the number of bath sites $n_b=6$ is used in the ED impurity solver.}
   \label{fig:edge}
   \end{center}
\end{figure}

To complete our discussion of the half-filled AKMH model, we show that the antichiral edge states persist at finite interaction strengths within the paramagnetic phase. Figure~\ref{fig:edge} shows the momentum-resolved spectral function $A_{y,\alpha}(\omega,k_x)$ for the cylindrical geometry illustrated in Fig. 1 in the main text, with $N_y=40$. The spectral function is averaged over the two inequivalent lattice sites within the unit cell along the $x$ direction. The horizontal white line at $\omega=0$ marks the Fermi energy. The results are obtained for $\lambda=0.1t$ and $U=2.0t$, for which the system remains in the paramagnetic phase. For each spin projection $\alpha$, copropagating edge modes connecting the two inequivalent Dirac points are visible at both the lower ($y=0$) and upper ($y=39$) zigzag edges. These edge-localized modes are absent from the representative bulk spectrum at $y=19$, where the compensating counterpropagating bulk states remain. The spectra of the two spin sectors are related by time-reversal symmetry, and their propagation directions are reversed. By incorporating dynamical electronic correlations beyond the noninteracting treatment of Ref.~\cite{Colomes2018}, our results demonstrate that the antichiral edge states remain robust at moderate interaction strengths and support their possible realization in the paramagnetic phase of TMD monolayers.

%